# Design and Synthesis of Ultra Low Energy Spin-Memristor Threshold Logic

Deliang Fan, Mrigank Sharad, Kaushik Roy

*Abstract*— **A threshold logic gate (TLG) performs weighted sum of multiple inputs and compares the sum with a threshold. We propose Spin-Memeristor Threshold Logic (SMTL) gates, which employ memristive cross-bar array (MCA) to perform current-mode summation of binary inputs, whereas, the low-voltage fast-switching spintronic threshold devices (STD) carry out the threshold operation in an energy efficient manner. Field programmable SMTL gate arrays can operate at a small terminal voltage of ~50mV, resulting in ultra-low power consumption in gates as well as programmable interconnect networks. We evaluate the performance of SMTL using threshold logic synthesis. Results for common benchmarks show that SMTL based programmable logic hardware can be more than 100× energy efficient than state of the art CMOS FPGA.**

## I. INTRODUCTION

IN recent years, several device solutions have been proposed for fabricating nano-scale programmable resistive elements, generally categorized under the term 'memristor' [1-3]. Of special interest are those that are amenable to integration with state of the art CMOS technology, like memristor based on Ag-Si filaments [2]. Such devices can be integrated into metallic cross-bars to obtain high-density memristive cross-bar arrays (MCA). Continuous range of resistance values obtainable in these devices can facilitate the design of multi-level, non-volatile memory [3, 4]. Application of the specific device characteristics of memristors in unconventional, computing schemes like neural networks [5, 6] and threshold logic (TL) [7-9], has been explored in recent years.

A threshold logic gate (TLG) operation essentially constitutes of summation of weighted inputs, followed by a threshold operation [10] (eq. 1). While a memristor array can be employed to perform current-mode analog summation of binary input voltage signals, the thresholding operation requires the application of a current comparator circuit. Such a comparison operation can be obtained using conventional analog circuits based on current mirrors [7] or voltage-comparators [8, 9]. However such analog CMOS circuits often consume significant power and area, thereby eschewing the energy and density benefits of nano-devices. Rather than depending upon analog CMOS circuits for implementing current comparison, it would be desirable to explore nano-devices that can directly provide such a current-mode thresholding characteristic.

Recent experiments on spin-torque devices have demonstrated high-speed switching of scaled nano-magnets with small current densities [11-14]. Such a phenomenon can be used to design compact and low-voltage current-mode spintronic switches and simultaneously provide energy-efficient current-to-voltage conversion. Application of such spin-torque switches in memory [28, 29], digital [15, 30], analog [16], and neuromorphic computing applications [17], have been explored earlier. Such nano-scale, spintronic devices inherently act as compact, ultra-low voltage and fast current-comparators and hence, can be highly suitable for memristor based TLG design.

In this work we propose spin-memristor threshold logic (SMTL) design using such spin-torque switches based on magnetic domain wall (DW) motion [13]. The magneto-metallic domain wall switch allows ultra-low voltage operation of memristive TLGs leading to low energy dissipation at the gate level. We name our proposed domain wall switch structure as spintronic threshold device (STD). It can facilitate ultra-low voltage current-mode interconnect for the design of fully programmable, large TL-blocks. This helps to achieve highly reduced energy dissipation in programmable interconnects. Notably, in CMOS-look up table (LUT) based conventional FPGAs, more than 90% of energy can be ascribed to programmable switches and interconnects [19]. Further, the STD being non-volatile magnetic switches inherently act as a latch and hence can facilitate fully pipelined connection of multiple TLG stages without the insertion of additional memory elements like flip-flops. This can provide high-performance and integration density for complex data processing blocks. The aforementioned factors combined together, lead to ultra-low energy consumption.

In this work, we also present a comprehensive methodology for SMTL design, synthesis and optimization and compare its performance with conventional CMOS FPGAs. The remainder of this paper is organized as follows. Section II discusses some of the previous work related to the design of memristor cross-bar array for weighted sum of threshold logic inputs. In section III, we introduce the spintronic threshold device for the sign function in TLG. Circuit design and optimization for SMTL are presented in section IV and section V respectively. Section VI presents the SMTL synthesis methodology. The performance and prospects of SMTL is discussed in section VII. Section VIII concludes the paper.

This work was supported in part by CSPIN, StarNet center, DARPA UPSIDE, SRC, Intel, and NSF.

The authors are with the School of Electrical and Computer Engineering, Purdue University, West Lafayette, IN, 47906, USA ( e-mail: dfan@purdue.edu; msharad@purdue.edu; kaushik@purdue.edu ).

## II. TLG First Stage Design using MCA

In this section we review the recent progress in memristive cross-bar array (MCA) design, programming and its application as the first stage of threshold logic computation.

A threshold logic operation shown in fig. 1a, can be expressed in the form of eq.1:

$$Y = sign(\sum(X_i W_i + b_i))\quad(1)$$

where, $X_i$'s are multiple binary inputs to a threshold gate, $W_i$'s are scalar weights with which the corresponding inputs are multiplied (or scaled) and $b_i$ is the bias for the $i^{th}$ gate. Note that, $W_i$ can be either positive or negative. Hence, depending upon the input combination (assuming unipolar values of inputs, i.e., 1 and 0) the summation can yield either a positive or a negative value, result of which is determined by the sign function (involving a comparison operation). The first stage of the threshold logic computation is the scaling and summation of the inputs, which can be implemented using a MCA, as shown in fig 1b. The detailed design and programming of MCA will be introduced in this section. The second stage of threshold logic computing is a 'sign' (in eq. 1, or threshold) function, which will be implemented using the proposed spintronic threshold device described in section III.

### A. Multi-level MCA

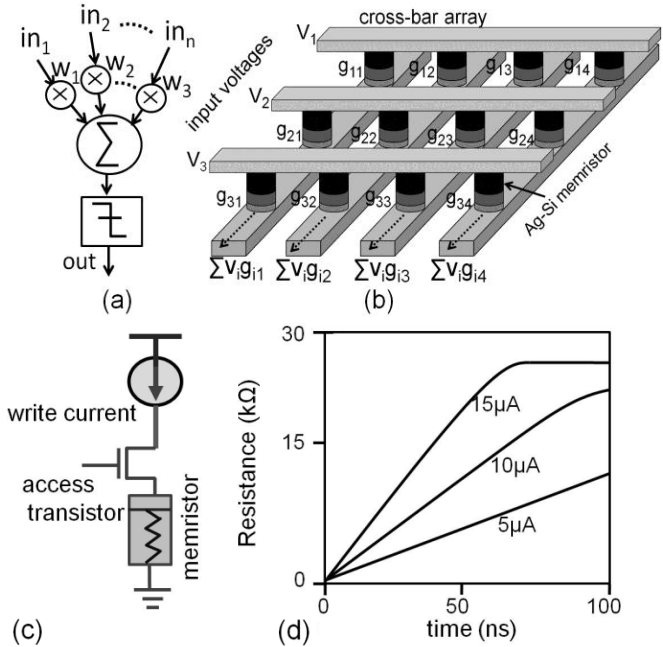

Fig. 1 (a) A Schematic representation of a threshold logic gate (TLG), (b) memristive cross-bar array, (c) A resistive memory cell with access transistors, (d) transient change in resistance for different magnitude of programming current.

Fig. 1b depicts a MCA that constitutes of memristors (Ag-Si) with conductivity $g_{ij}$, interconnecting two sets of metal bars ($i^{th}$ horizontal bar and $j^{th}$ in-plane bar). High precision, multi-level write techniques for isolated memristors have been proposed and demonstrated in literatures that can achieve more than 8-bit write-accuracy [3, 4]. However, for threshold logic design the bit-precision requirement can be significantly less (less than 4-bit, shown in fig. 13b). In a cross-bar array, consisting of large number of memristors, write-voltage applied across two cross-connected bars for programming the interconnecting memristor also results in sneak current paths through neighboring devices. This disturbs the state of unselected memristors. To overcome the sneak path problem, application of access transistors (fig. 1c), and diodes have been proposed in literature that facilitate selective and disturb free write operations [24]. Methods for programming memristors without access transistors have also been suggested, but using such techniques only a single device in an array can be programmed at a time [25]. Such schemes can be applicable only if programming speed is not a major concern.

Fig.2a depicts a possible array-level schematic of multi-level writing scheme for memristors, using adjustable pulse width [4]. The memristor cells to be written are selected by choosing the corresponding set of the word line, the source-line and the bit line. For infrequent write operations, a single write unit can be shared among large number of rows, as shown in fig. 2a. However, for maximum write-speed, each row can have a dedicated programming cell. This would allow writing of one column at a time, by selecting a particular world line. In order to accomplish the write operation, a constant current can be injected into the selected cell and the voltage developed on the source line is compared with a comparator threshold. The threshold, in turn, is set proportional to the target resistance, by using a compact switched capacitor digital to analog converter (DAC). The current source is disconnected as soon as the accessed memristor acquires the target resistance value. As shown in fig.1d, lower value of write current results in slower ramp in the resistance value and hence, can allow more precise tuning. Analytical model for memristor have been used for simulation in this work [4]. Experimentally it has been observed that memristive devices (including Ag-Si) do exhibit a finite write threshold for an applied current/voltage, below which there is negligible change resistance [26]. As described in the following sections, application of spin-based current comparator in MCA facilitates ultra-low voltage (and hence low current) operation of the memristors for computing and hence, can achieve reduced read-disturb for the array.

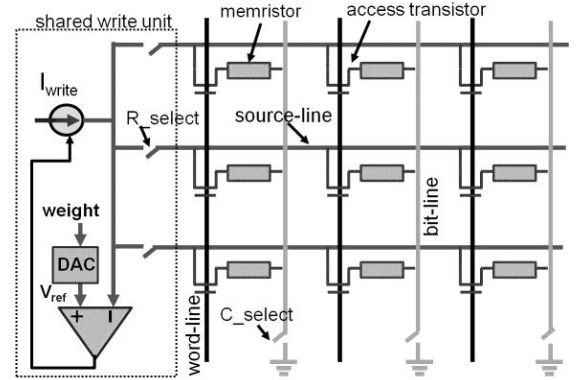

Fig. 2a: A resistive memory array with multi-level programming periphery.

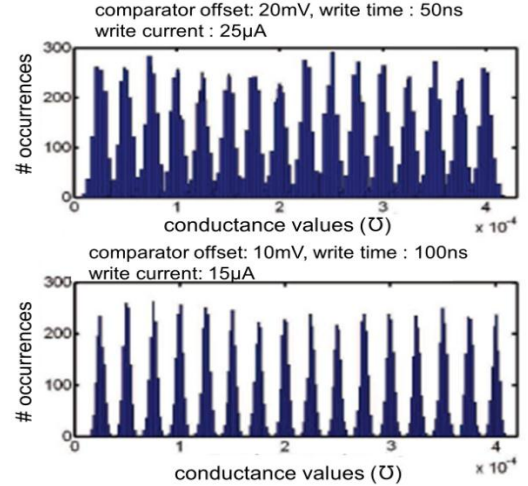

Fig. 2b: Simulation results for feed-back-based write show that higher write precision can be obtained by employing higher resolution comparator and longer write-time. These trends have been obtained using analytical model for memristors [4].

The write-precision in the method described above, is mainly limited by the random offset of the comparator, inaccuracy in the current source and the DAC. Larger accuracy would entail higher design-complexity for these blocks and lower write-speed (fig. 2b).

### B. Threshold Logic Computation Using MCA

For a TLG, the scaling and summation operations can be implemented using a MCA, as shown in fig 1b. If we assume that the outward terminals of the in-plane bars are connected to ground potential, for a given set of binary voltage inputs-$V_i$, the resulting current flowing out of the in-plane bars can be visualized as the dot product of the inputs voltages and the conductance values [3, 10].

The above principle can be exploited in realizing current-mode analog scaling (multiplication) and summation that corresponds to the first-stage operation of a TLG. Several authors have proposed the design of hybrid TLG hardware based on memristive cross-bar arrays and analog CMOS circuits, where analog circuits are employed to perform the second stage operation of the TLG, namely, thresholding [7-9]. For instance, application of analog current-mirrors have been proposed for impelementing memristor-based hybrid TLG's in [7]. However such a design requires additional interconnect networks to realize fully programmable logic modules. Notably, energy consumption of interconnects dominate the total power budget of an FPGA [19]. Authors in [8, 9] applied CMOS voltage comparators for realizing the thresholding operation for memristor-based TLGs. Application of analog amplifiers and comparators may lead to significant energy consumption. Authors in [10] recently demonstrated the use of a simple CMOS latch for thresholding operation. Such a scheme would need large voltage inputs (resulting in large current) to the memristors, so that a digital latch can directly sense the voltage-mode output of a TLG. This would result in power hungry TLG blocks that may not be suitable for large-scale integration.

Thus, although memristors can provide an efficient mapping of the first stage operation of a TLG (namely current-mode scaling/multiplication and summation), the second operation, namely, the current-mode thresholding, does not have a likewise 'matching' device. The above mentioned inefficiencies could be eliminated if an alternate device structure could be found that could perform the current-mode thresholding operation in an energy-efficient way. In the next section we present a spin-torque based device that can be ideally suitable for this purpose.

### III. TLG SECOND STAGE DESIGN USING SPINTRONIC THRESHOLD DEVICE

In this section we present the spintronic threshold devices (STD), based on magnetic domain wall, suitable for the design of energy efficient Spin-Memristor Threshold Logic (SMTL). This STD design will serve as the second stage of threshold logic computing, which is a thresholding ('sign') function in eq. 1.

The device structure for the STD is shown in fig. 3. It constitutes of a thin and short ($20 \times 40 \times 3$ nm$^3$) nano-magnet domain, d2 connecting two anti-parallel nano-magnet domains of fixed polarity, d1 and d3. Domain-1 forms the input port, whereas, domain-3 is grounded. Spin-polarity of the free-layer (d2) can be written parallel to d1 or d3 by injecting a small current along it from d1 to d3 and vice-versa [15-17]. Thus, the STD can detect the polarity of the current flow at its input node. Note, STD acts as an ultra low voltage and compact current-comparator that can be employed in the design of current-mode threshold logic.

The resolution of the device, i.e. the minimum-current magnitude required to switch the free layer, is determined by the critical current density for DW motion. Several recent experiments have achieved sub-nanosecond domain wall motion, with current density of ~$10^6$ A/cm$^2$ [11]. Magnetic

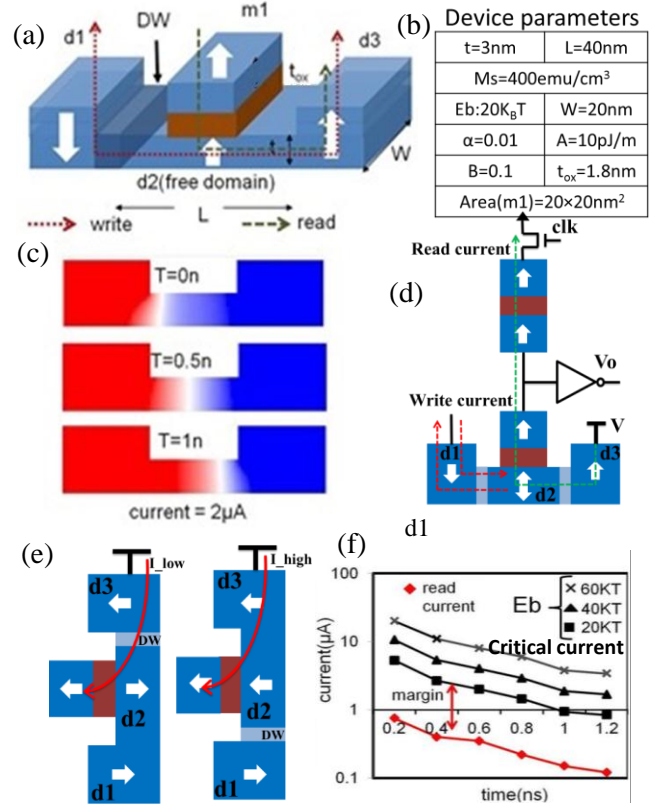

Fig. 3: (a) Device structure for STD (b) Device parameters table. (c)Transient micro-magnetic simulation plots. Read color represents the 'down spin' corresponding to d1. Blue color represents the 'up spin' in d3. White color is the magnetic domain wall. (d) STD state sense circuit (e) read current for different d2 state (f) read current margin to critical current

domain with perpendicular magnetic anisotropy can provide scaled device dimensions (thickness ~3nm and width <50nm) as well as relatively lower critical current density [12-14]. More recently, application of spin-orbital coupling has been explored for reducing the required current for a given speed of domain wall motion by an order of magnitude [14]. These device optimizations can be used to engineer current thresholds of the order of ~2μA for 1ns switching. Fig. 3c shows the transient micro-magnetic simulation plots for the proposed STD design using Object Oriented Micro-Magnetic Framework (OOMMF, [31]) when supplied with a 2 μA current. It can be seen the magnetic domain wall moves from the left free domain boundary to the right boundary within 1 ns. We will analyze the effect of STD resolution on the energy efficiency of SMTL in section VII.

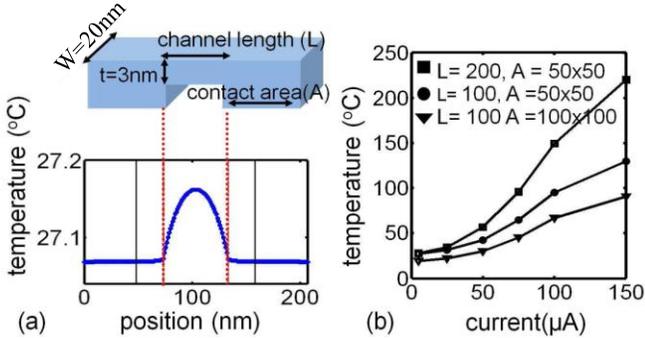

Fig. 4 (a) COMSOL simulation for temperature rise in the STD device for different device dimensions, (b) plot showing temperature profile along the device for a small input current

A magnetic tunnel junction (MTJ) formed between a fixed polarity magnet m1 and the free-domain d2 is used to read the state of d2. The effective resistance of the MTJ is smaller when m1 and d2 have the same spin-polarity and vice-versa (design parameter in fig. 3b). The ratio of the two resistances is defined in terms of tunnel magneto-resistance ratio (TMR). The STD-MTJ forms a voltage divider with a fixed reference MTJ, as shown in fig. 3d. A TMR of ~400% can provide a voltage swing close to VDD/2 that can be detected using a simple CMOS inverter. Static current in the voltage divider can be minimized for a given operation speed by increasing the MTJ oxide thickness. For 500MHz clock frequency, the oxide thickness was determined to be ~1.8nm that resulted in a total power dissipation of ~0.15μW for the sensing unit (including the clocking power), for a supply voltage of 0.6V.

Note that in the detection circuit, the terminal d3 of the STD is connected to Vdd. Hence, the transient evaluation current flows from d3 to d2 as shown in fig. 3e. The current required for the DW motion increases proportional to the switching speed. Since the transient read current flows only for a short duration and the magnitude is lower than the critical current to move the DW, it does not disturb the state of d2. The read margin can be seen in fig. 3f. Apart from device scaling, the STD critical current can also be lowered by manipulating other device parameters, like the anisotropy energy (Eb) of the magnet (fig. 3f).

The reliability of a magnetic domain wall motion device is excellent. The velocity of domain wall and critical current are not sensitive to the external magnetic field or temperature [27]. 10-year retention time at 150ºC and $1 \times 10^{14}$-times write endurance for the Co/Ni wire are also reported in [27]. Analyzing the heating effect on the magneto-metallic STD is critical for reliability assessment of the device. The effect of Joule heating in the STD was simulated using finite-element simulation through COMSOL [28]. The thin and short central free domain of the device is the most critical portion with respect to current driven heating (fig. 4a). Fig. 4b shows that the heating in the device can be reduced by choosing larger contact area of the two fixed domains. Also, shorter free domain results in smaller heating. Thus, the current handling capacity of the device can be increased by appropriate structural optimization.

In general, fig. 3d circuit forms the 'sign' function in eq. 1. The STD works as a current-comparator and its input is the output current of the first stage MCA. If the input current to STD is larger than the critical current, the output of the inverter in fig. 3d is high (vice versa). Next, we describe circuit design for combining the MCA and STD to implement threshold logic array design.

## IV. DESIGN OF SMTL ARRAY

Fig. 5a and 5b show two threshold logic networks (TLN)

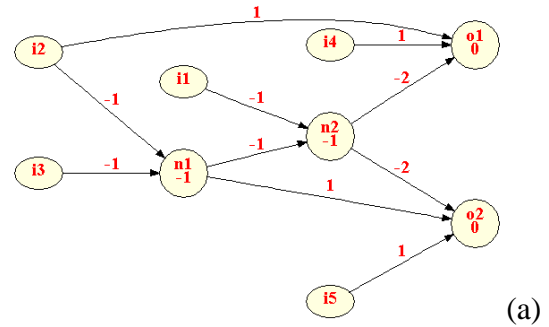

Fig. 5a: synthesized ISCAS85 benchmark C17 threshold logic network. Each circle represents one threshold logic gate. The connections between each TLG are the fan-ins and fan-outs. The node without fan-ins is the input node. The node without fan-outs is the output node. The weights are labelled along the connections. i1-i5 are the input nodes, n1 and n2 are TLGs, o1 and o2 are the output nodes. The bias values are labelled inside of the TLGs. The nodes in the same column are in the same stage

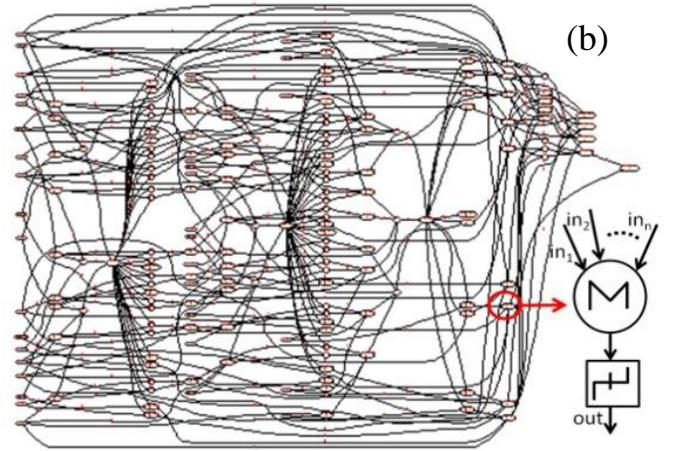

Fig. 5b: synthesized ISCAS85 benchmark-C432 (27-channel interrupt controller) threshold logic network. The weight range is shown in fig. 13. This synthesized threshold logic network consists of 15 stages, while each stage is comprised of Ni threshold logic gates. The maximum fun-in for each TLG is 4.

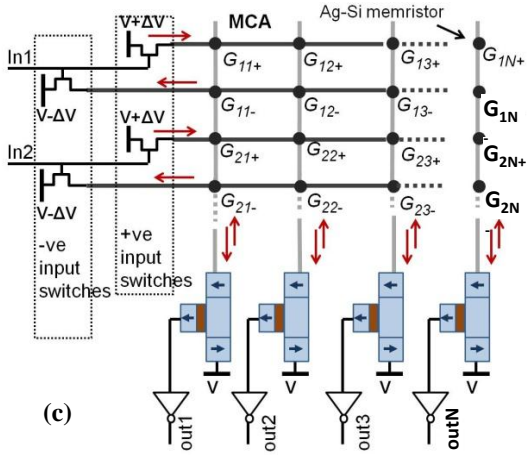

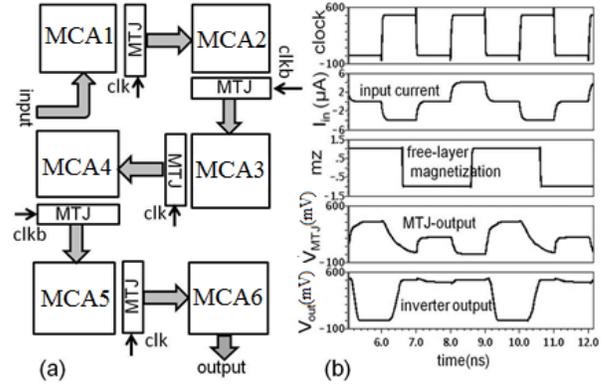

Fig. 6: (a) 2-phase pipelined MCA blocks for large-scale logic design, (b) transient simulation plots for a single TLG.

Fig. 5c: Circuit of one single threshold logic stage using MCA and STD

for an ISCAS85 benchmark, C-17 and C-432 [7], obtained using the threshold logic synthesis (TELS) technique presented in [18]. The meaning of the symbols is explained in the caption. C-17 is a simple TLN, while C-432 is a much larger scale TLN. In order to show our design is compatible to large scale TLN mapping, we will use C432 as a design example in the following paper.

TLN constitutes of a network of TLGs which can be divided into multiple stages. Each circle in the plot represents one TLG and the TLG in the same columns will be mapped to the same MCA stage. The connections between the TLGs are implemented by the MCA described in section II, whereas the conductance of memristor corresponds to the synthesized weights. In such a multi-stage logic scheme, each MCA stage would comprise a number of TLGs receiving inputs from its previous stage and communicating their outputs to the next stage. Let us consider the design of such a stage using MCA and the STD device.

Fig. 5c shows the circuit realization of a single MCA stage that contains N number of TLGs based on STD. Each stage has a maximum of M inputs (which can be set as a parameter during the MCA mapping), and N STDs, forming the N TLGs. The $i^{th}$ input to the MCA may connect to the $j^{th}$ STD (i.e. $j^{th}$ TLG) with either a positive, negative or zero weight. This is achieved by programming either of $G_{ij+}$ or $G_{ij-}$ to the corresponding weight value. For zero weight (i.e. no connectivity), both $G_{ij+}$ and $G_{ij-}$ are driven to high-resistance off-state. The input-signal to MCA is received through PMOS transistors with source terminals connected to a potential $V+\Delta V$ (for positive weights) and $V-\Delta V$ (for negative weights) where $\Delta V$ can be less than ~50mV. These input transistors act as deep-triode region current sources (DTCS) [16]. The STD is connected to a DC supply V. This effectively clamps the potential of all the vertical metal bars in fig. 5c to the same potential (due to small resistance of the magneto-metallic STD). The static current employed in computing therefore flows across a small terminal voltage of $\Delta V$, resulting in small static power consumption. Moreover, the dynamic power dissipation on the metallic interconnects forming the programmable cross-bar is also largely reduced due to ultra-small voltage swing. The direction of current flow at the input of a STD, and hence the output of a TLG, would depend upon the input data and the corresponding weights (determined by the programmed memristor conductance). Note that, the resistance values for the memristors can be chosen large enough to avoid inaccuracy due to resistive voltage division between the DTCS transistors and the memristors in a given row. The output of the MTJ-based detection circuit associated with each TLG, in turn, drives a corresponding DTCS transistor that communicates the outputs of the TLGs to the next stage.

Due to the non-volatility of the STD, the MCA design described above can be extended to realize a 2-phase pipelined architecture composed of large number of such hybrid arrays without inserting the CMOS latches, as shown in fig.6a. In such a design, consecutive MCAs operate with complementary clock phases. For instance, in fig. 6a, when the clock is high, MCA1 is driving MCA2, and MCA3 is driving MCA-4. When the clock goes low, the driver and driven MCAs exchange roles. The exemplary simulation plots for a single TLG is shown in fig. 6b.

Next we discuss optimal pipelining and partitioning scheme for the mapping of large logic blocks on to the SMTL array.

V. OPTIMAL PIPELINING AND PARTITIONING OF SMTL ARRAYS FOR LOGIC MAPPING

A. Pipeline-Optimization:

As mentioned earlier, each STD acts as a non-volatile latch and hence, a multi-stage MCA can be pipelined without insertion of additional CMOS latches. However, logic paths in the threshold logic network (TLN) of a generic logic block (like for C432 shown in fig. 5a) may be unequal. Hence 'buffer-nodes' need to be inserted to make them equal and to facilitate fine-grained pipelining. The number of buffers needed depends upon the granularity of pipelining. In case, each MCA stage is pipelined, the number of buffers is the maximum. Fully-pipelined TLN for C432 is shown in fig. 7a. In such a TLN, each stage is mapped into a separate MCA stage. For a given switching-speed of the STD, this configuration yields maximum throughput. However, the total energy consumption also depends upon the total number of TLG nodes.

Combining two MCA stages to form a single pipelined stage (fig. 7b) reduces throughput by half, however the total number of nodes for most benchmarks was found to reduce by a larger factor, which leads to reduced energy consumption.

Note that, despite using multiple MCA layers per pipeline-stage, the same throughput can be maintained by increasing the current injection, i.e., the switching-speed of the STD.

Fig. 8 shows the power consumption for C432 for different number of MCA levels (note, single MCA level for a pipelined stage implies maximum pipeline granularity) in a single pipelined stage. The power component due to the detection unit ('Power_det' due to MTJ-voltage divider, clock and inverter) reduces with reducing pipeline granularity, because of reduction in total number of TLG nodes in the resulting TLN (fig. 8b). However, to maintain the same throughput, larger currents need to be supplied by the DTCS transistors, which lead to increase in static power consumption in the MCA ('Power_MCA' in fig.8a). For most ISCAS85 benchmarks a pipelined stage with 2-MCA levels yielded optimal results (fig. 8b).

### B. Partition and Interconnects

So far we assumed that each stage of the pipelined TLN is assigned to a single large-dimension MCA. In such a design no additional interconnect network is required, as, the outputs of the $N^{th}$ MCA stage can directly connect to the inputs of the $(N+1)^{th}$ MCA stage using the scheme shown in fig. 7. Due to the absence of additional interconnect power dissipation, this leads to the minimum energy solution (fig. 10a). However, in this case, the MCAs have sparse connectivity (due to having large number of inputs but each input connecting to only few outputs, determined by the fan-in limitation) due to which the overall area efficiency is significantly sacrificed, as shown in fig. 10b. To reduce the overall area, each pipeline stage can be divided into multiple smaller dimension sub-arrays (Ai's shown in fig. 7b and an enlarged version in fig. 9a). In this case, some of the inter-layer connections can still be directly routed to the next stage (fig. 9a). However, some others (between nodes that are not located on directly opposite MCAs) need to be routed through an additional routing network. Such a design-scheme is shown in fig. 9b. For reducing MCA dimensions (implying the use of large number of smaller MCA modules in a single stage), the usage of the interconnect network increases. This also necessitates larger and longer interconnect array, leading to larger parasitic resistance drops along the current-signal paths, mandating the use of larger voltage. As a result, energy component due to interconnect increases. Fig. 10 shows the tradeoff between area and power of SMTL with respect to the size of the sub-MCA array size. A design choice can be made based on priority.

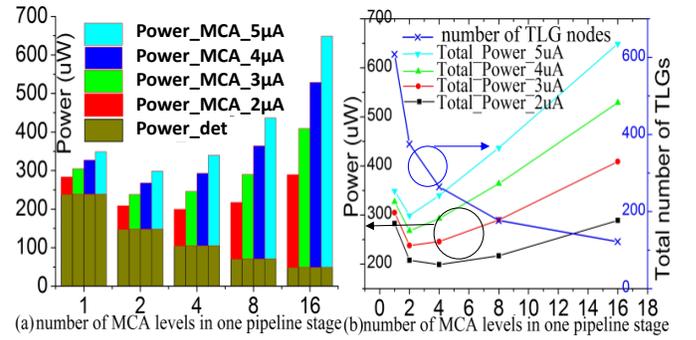

Fig. 8: (a) Power consumption of different pipeline configuration (b) tradeoff between power and area. 'Power_MCA_5uA' represents the power of memristor cross-bar array when the DTCS current is 5uA. 'Power_det' is the power of detection module including MTJ-voltage divider, clock and inverter

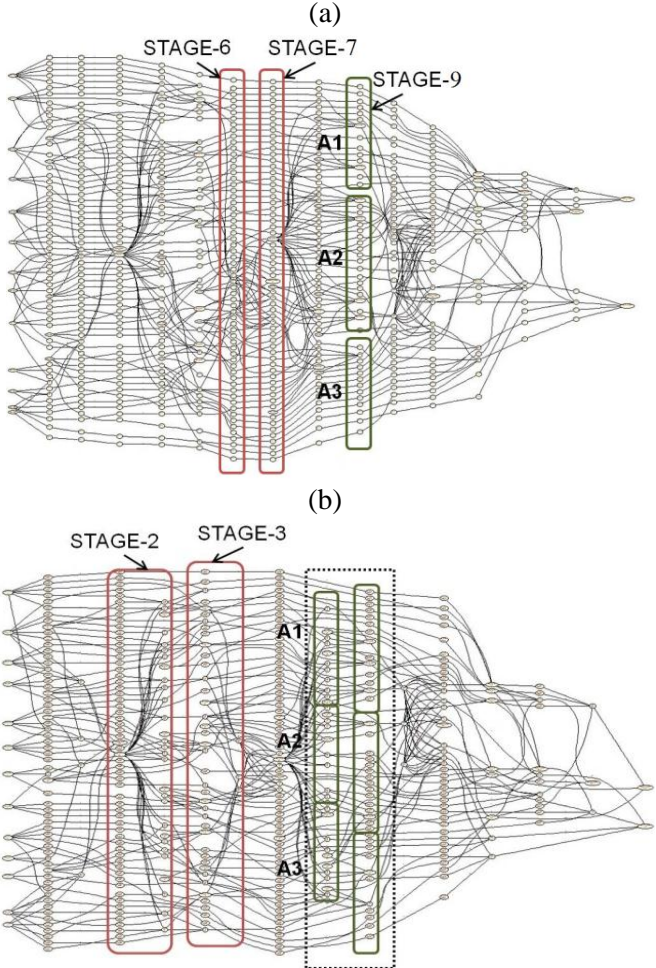

Fig. 7: synthesized C432 pipelined threshold logic network. (a) Fully pipelined architecture (b) two TLG stages combined with one pipeline stage. Each circle represents one TLG and the TLGs in the same column are in the same stage.

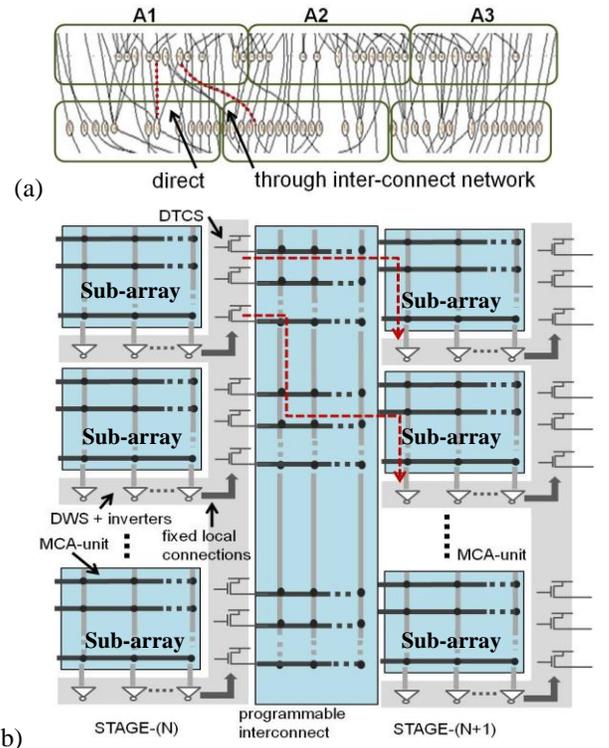

Fig. 9: (a) enlarged green square part of fig 7b (b) SMTL network partition architecture

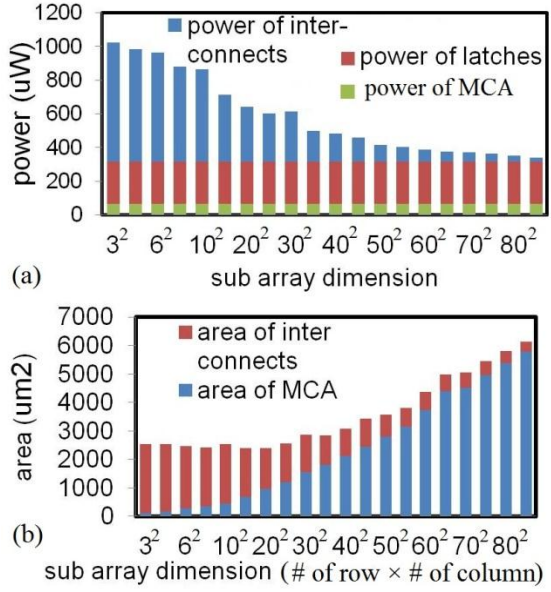

Fig. 10: relationship between (a) power, (b) area and sub-array dimension, (larger dimension implies lower number of sub-arrays needed)

## VI. SIMULATION AND SYNTHESIS ALGORITHM

In this section we discuss the synthesis scheme used in this work to assess the performance of SMTL.

Fig. 11 shows the high level overview of the SMTL-synthesis and hardware-mapping methodology employed in this work. We employed threshold logic synthesis (TELS) algorithm proposed in [18] to do the initial synthesis, which reads a logic description and generates the functionally equivalent threshold network. Some important parameters like the fan-in restriction of TLGs and defect-tolerance in the weights can be preset as parameters.

The SMTL mapping algorithm proposed and implemented in this paper, shown in fig. 12, reads the synthesized TLG network and maps it to SMTL hardware. The tool first reorders the positions of TLGs in each stage so as to minimize the use of the interconnect network. This is achieved by placing the TLGs in the sub-arrays such that the use of direct links between face-to-face MCAs (as depicted in fig. 9a) is maximized. Next, if the number of nodes in the current stage exceeds the restriction (number of MCA in a given stage times MCA size), one or more nodes are moved to next stage. This is done in a way that minimizes the number of intermediate buffers. The nodes without fan-out to next one stage are selected with highest priority, following which, the nodes with minimum fan-in's are shifted.

Some of the layers in the SMTL netlist may have very small number of nodes, for which, the use of a separate MCA unit may be wasteful. In TELS such nodes are incorporated in the MCA units corresponding to the previous stage, through the provision of a small numbers of programmable backward connections (from output of an MCA back to its input).

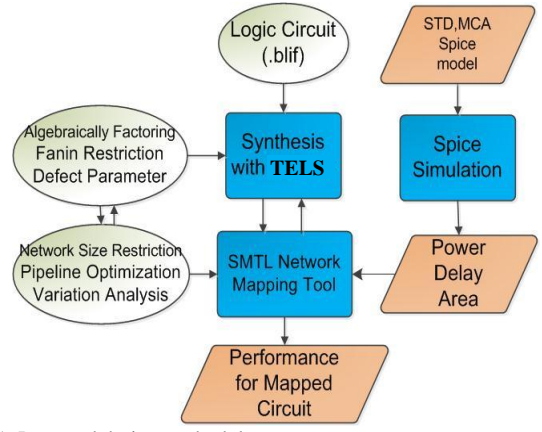

Fig. 11: Proposed design methodology

The fan-out number of some nodes can be very large. Such TLGs communicate evenly to all the MCAs in the next level, making heavy use of the interconnect network. Such high-loading can lead to significant voltage division between the DTCS source and the receiving memristors, leading to significant lowering of the input voltage and the current for the loads. A simple way to address this issue is to split the large fan-out nodes into multiple smaller nodes.

Larger TLG fan-in generates denser SMTL network with smaller number of TLG nodes. This can provide larger area and energy efficiency. However, simulations show that larger fan-in restriction leads to reduced variation tolerance for memristor values, as seen in fig. 13b. In this plot, variation tolerance is defined as the standard-deviation ($\sigma$) value for which total $10^5$ test vector simulation gave zero errors. The variation tolerance increases for lower fan-in restriction, but the use of lower fan-in TLGs results in larger number of nodes, leading to increase in overall area (fig. 13). In this work we choose the fan-in restriction to be 4 (leading to a variation tolerance of ~9%, as shown in fig. 13). There are only 6 different levels of memristor conductance needed for mapping the TLG weighs, therefore the programming bit resolution for memristor is 3 bit. Note that, in this work we have assumed that the memristor programming thresholds are large enough, such that passing small computing currents (few µA) does not significantly disturb their state [2].

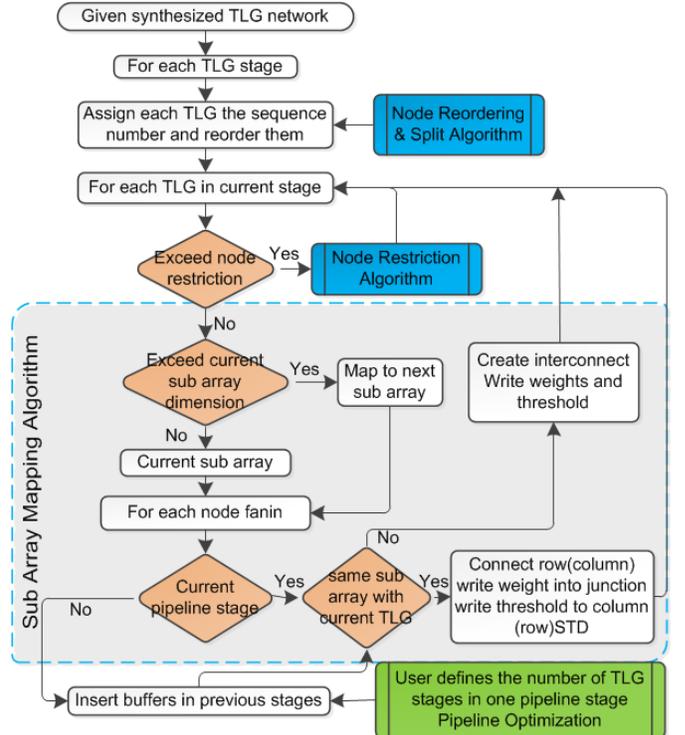

Fig. 12: SMTL network mapping algorithm

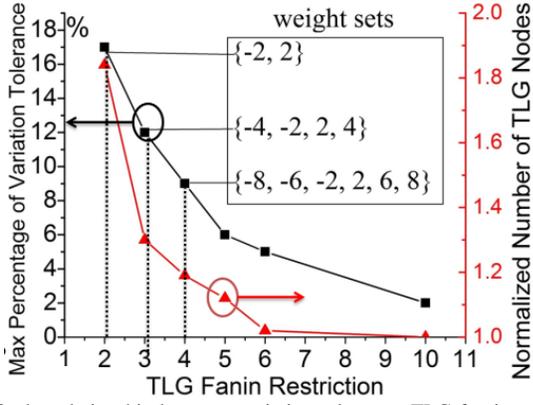

Fig. 13: the relationship between variation tolerance, TLG fan-in restriction and number of TLGs

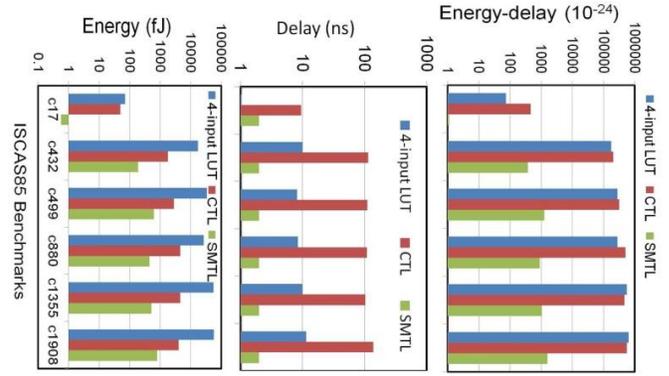

Fig. 14: (a) Computation energy, (b) delay and (c) energy-delay product of SMTL compared with 4-input LUT based FPGA [7] and CTL [7] for ISCAS85 benchmarks. ( CTL: a CMOS based implementation style for TLG [7]).

Next we discuss the performance of SMTL and compare it with conventional CMOS programmable logic based on CMOS LUTs.

## VII. PERFORMANCE AND PROSPECTS

In the conventional FPGA based TLG design, the interconnect power is the bottleneck of the total power consumption. Note that more than 90% of energy can be ascribed to programmable switches and interconnects [19]. The reason is the fact that the FPGA interconnect circuit has an extremely low utilization rate (~12%) for purpose of programmability. The energy and delay of 4-input LUT based FPGA for ISCAS85 benchmark using 45 nm technology is shown in fig. 14. While in our proposed SMTL design, the energy efficiency mainly comes from four aspects. **1)**: The interconnect energy dissipation in the metallic cross-bars as well as the interconnect network is drastically lowered due to ultra-low voltage (~50mV), current-mode signaling between the MCA layers, which comes from low voltage, low current operation of spin-torque based threshold logic gates. The STD device can sense and compare the ultra-low current (few µA) enabling ultra-low voltage biasing of the MCA and hence, low voltage operation of the threshold gates. As a result the static power consumption, due to direct current paths, is largely reduced. Note that in the SMTL design, memristors play the dual role of computing elements as well as programmable interconnects. This can be contrasted with earlier approaches where memristors were employed only as programmable interconnects [21] or only as computing elements [7]. **2)**: In our proposed threshold logic network design, the output inverters of a particular MCA layer drives only the DTCS transistors that in-turn supply current to the next MCA stage. The MCA itself is operated across a small terminal voltage $\Delta V$, thereby reducing the $CV^2 f$ dynamic power consumption in large number of programmable interconnects. Such low-voltage operation of the MCA can also significantly reduce the distrub rate of the programmed memristors and can enhance the retention time of the hardware. **3)**: The STD achieves energy efficient current-to voltage conversion with the help of MTJ-based voltage divider. This eliminates the need of analog trans-impedance circuits based on current mirrors and amplifier, leading to high energy and area efficiency. **4):** Due to the non-volatility of STD, the proposed SMTL design can be extended to realize a pipelined architecture without inserting the CMOS latches. The throughput of the design is determined by a single stage delay. This delay in turn, is limited by the switching speed of the STD device. As discussed earlier, larger current per input can be used to increase the STD-switching speed. Domain wall velocities of more than 400m/s has been demonstrated in literature [22], hence, for a 40nm long free domain more than 1GHz processing speed may be acheivable. In this work a clock frequecny of 500MHz has been used, corresponding to STD switching time of 1ns. Recently application of spin-hall effect has been explored for bringing large reduction in domain-wall current thresholds [13]. Such phenomena can be exploited in improving the resolution of scaled STD devices.

Fig. 14a compares the computation energy of the proposed SMTL design with that of 4-input lookup table (LUT) based FPGA and with capacitive threshold logic (CTL, a CMOS based implementation style for TLG [7]). It shows about two orders of magnitude lower computing energy for the proposed design as compared to the LUT based FPGA TLG. SMTL also shows much smaller delay compared with LUT and CTL, as shown in fig. 14b. Results in fig. 14c show around three orders of magnitude lower energy-delay product as compared to both the CMOS based schemes.

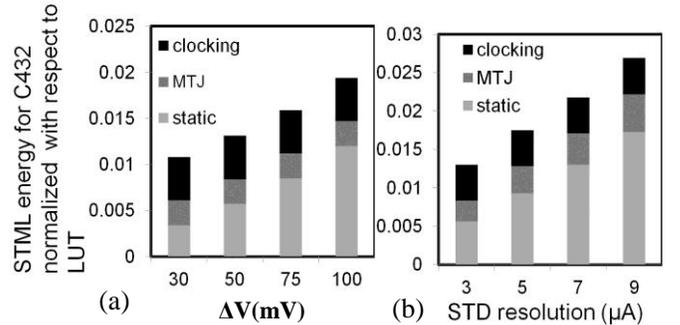

Fig. 15: SMTL energy for C432 normalized with respect to 4-input LUT for the case of (a) increasing $\Delta V$, (b) increasing STD threshold for a fixed $\Delta V$ of 50mV ;LUT delay is ~10ns

TABLE I
DESIGN PARAMETERS

| | | | |
|---|---|---|---|
| free-domain size | $3\times20\times40 nm^3$ | MTJ-$t_{ox}$ | 1.8nm |
| Ms | 400 emu/cm$^3$ | $R_{MTJ}$ (parallel) | 300kΩ |
| $K_{u2}V$ | $20K_BT$ | MTJ-TMR | 400% |
| β (nonadiabatic const.) | 0.1 | MTJ-area (nm$^2$) | 20x20 |
| α (damping coeff.) | 0.01 | Memristor values (Ω) | 50k to 1M |
| $I_{threshold}$ for STD | 2μA | ΔV | 50mV |
| V | 0.6V | CMOS tech. | 45nm |

The energy efficiency of the proposed design is dependent on two critical design parametrs. First, the minimum achievable ΔV in such a hybrid circuit. Figure. 15a shows that increasing ΔV increases the static power consumption due to current-mode computing in MCAs (strength of DTCS transitors is reduced to keep the current drive constant). The second important parameter is the resolution of the STD device. As mentioned earlier, a poor resolution would require larger current per-input for a TLG. Corresponding results are shown in fig. 15b, showing almost linear increase in computation energy with reducing resolution.

Integration of Ag-Si memristors with CMOS has been demonstrated in recent years [2, 3]. The same is true with magnetic domain wall based memory cells [13, 23, 27]. However, integrating two novel technologies with CMOS to realize the proposed SMTL scheme can be significantly more challenging, especially when scaled dimensions of STD devices, such as used in this work, is targetted. However, the possiblity of large energy benefits of the proposed design can be a motivating factor.

Some critical design parameters used in this work are given in table I. The device characteristics for STD was obtained using the micromagnetic simulation framework for domain wall magnet presented in [23]. Behavioral model based on statistical characteristics of the device were used in SPICE simulation to assess the system level functionality.

## VIII. CONCLUSION

Spintronic threshold device can be combined with CMOS compatible Ag-Si memristors for designing ultra low energy Spin-Memristor Threshold Logic (SMTL). Such a hardware can achieve more than 100× improvement in energy and 1000× improvement in energy-delay product, as compared to state of the art CMOS FPGA based TLG, due to low voltage, low current computing facilitated by a spin-torque device.

ACKNOWLEDGMENT

This research was funded in part by DARPA, MARCO StarNet, SRC, and NSF.

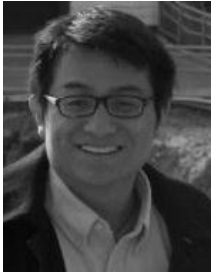

**Deliang Fan** received his B.S. degree in Electronic Information Engineering from Zhejiang University, China, in 2010 and M.S. degree in Electrical and Computer Engineering from Purdue University, IN, USA, in 2012. Currently he is a graduate research assistant of Professor Kaushik Roy and pursuing Ph.D. degree in Electrical and Computer Engineering at Purdue University.

His primary research interest lies in cross-layer (algorithm/architecture/circuit) co-design for low-power Boolean, non-Boolean and neuromorphic computation using emerging technologies like spin transfer torque devices. His past research interests include cross-layer digital system optimization and imperfection-resilient scalable digital signal processing algorithms and architectures using significance driven computation.

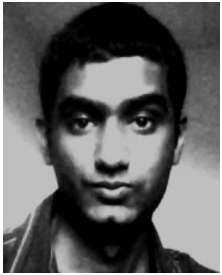

**Mrigank Sharad** received the Bachelor's and Master's degree in electronics and electrical communication engineering from IIT Kharagpur, India, in 2010, where he specialized in Microelectronics and VLSI Design. Currently he is working toward the Ph.D. degree in electrical and computer engineering at Purdue University.

His primary research interests include low-power digital/mixed-signal circuit design. His current research is focused on device-circuit co-design for low power logic and memory, with emphasis on exploration of post-CMOS technologies like, spin-devices. He has pioneered the concept of spin-CMOS hybrid design for ultra-low power neuromorphic computation architectures. He has also worked on application of spin-torque devices in approximate computing hardware and memory design.

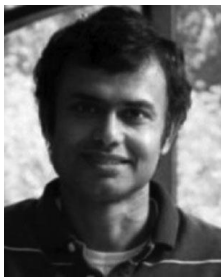

**Kaushik Roy** (F'02) received B.Tech. degree in electronics and electrical communications engineering from the Indian Institute of Technology, Kharagpur, India, and Ph.D. degree from the electrical and computer engineering department of the University of Illinois at Urbana-Champaign in 1990. He was with the Semiconductor Process and Design Center of Texas Instruments, Dallas, where he worked on FPGA architecture development and low-power circuit design. He joined the electrical and computer engineering faculty at Purdue University, West Lafayette, IN, in 1993, where he is currently Edward G. Tiedemann Jr. Distinguished Professor. His research interests include spintronics, device-circuit co-design for nano-scale Silicon and non-Silicon technologies, low-power electronics for portable computing and wireless communications, and new computing models enabled by emerging technologies. Dr. Roy has published more than 600 papers in refereed journals and conferences, holds 15 patents, graduated 60 PhD students, and is co-author of two books on Low Power CMOS VLSI Design (John Wiley & McGraw Hill).

Dr. Roy received the National Science Foundation Career Development Award in 1995, IBM faculty partnership award, ATT/Lucent Foundation award, 2005 SRC Technical Excellence Award, SRC Inventors Award, Purdue College of Engineering Research Excellence Award, Humboldt Research Award in 2010, 2010 IEEE Circuits and Systems Society Technical Achievement Award, Distinguished Alumnus Award from Indian Institute of Technology (IIT), Kharagpur, Fulbright-Nehru Distinguished Chair, and best paper awards at 1997 International Test Conference, IEEE 2000 International Symposium on Quality of IC Design, 2003 IEEE Latin American Test Workshop, 2003 IEEE Nano, 2004 IEEE International Conference on Computer Design, 2006 IEEE/ACM International Symposium on Low Power Electronics & Design, and 2005 IEEE Circuits and system society Outstanding Young Author Award (Chris Kim), 2006 IEEE Transactions on VLSI Systems best paper award, 2012 ACM/IEEE International Symposium on Low Power Electronics and Design best paper award, 2013 IEEE Transactions on VLSI Best paper award. Dr. Roy was a Purdue University Faculty Scholar (1998-2003). He was a Research Visionary Board Member of Motorola Labs (2002) and held the M.K. Gandhi Distinguished Visiting faculty at Indian Institute of Technology (Bombay). He has been in the editorial board of IEEE Design and Test, IEEE Transactions on Circuits and Systems, IEEE Transactions on VLSI Systems, and IEEE Transactions on Electron Devices. He was Guest Editor for Special Issue on Low-Power VLSI in the IEEE Design and Test (1994) and IEEE Transactions on VLSI Systems (June 2000), IEE Proceedings -- Computers and Digital Techniques (July 2002), and IEEE Journal on Emerging and Selected Topics in Circuits and Systems (2011). Dr. Roy is a fellow of IEEE.